\documentclass[10pt,conference,letterpaper]{IEEEtran}
\usepackage{times,amsmath,epsfig}
\title{Incorporating Integrity Constraints in Uncertain Databases}
\author{%
{Naveen Ashish{\small $~^{\ 1}$}, Sharad Mehrotra{\small $~^{ 2}$}, Pouria Pirzadeh{\small $~^{\ 3}$} }%
\vspace{1.6mm}\\
\fontsize{10}{10}\selectfont\itshape
$~^{\ }$Calit2 and ICS Department, UC Irvine\\
Irvine CA 92697 USA\\
\fontsize{9}{9}\selectfont\ttfamily\upshape
$~^{1}$ashish@ics.uci.edu\\
$~^{2}$sharad@ics.uci.edu%
}

\usepackage{amsmath}
\usepackage{comment}
\usepackage{graphicx}
\usepackage{cite}
\usepackage{url}
\usepackage{epsfig}
\usepackage{varioref}
\usepackage{bbm}

\begin{document}

\maketitle

\begin{abstract}
We develop an approach to incorporate additional knowledge, in the form of
general-purpose integrity constraints (ICs), to reduce uncertainty in
probabilistic databases. While incorporating ICs improves data quality (and hence quality of answers
to a query), it significantly complicates query processing. To overcome the additional
complexity, we develop an approach to map an  uncertain relation $U$ with ICs to  another
uncertain relation $U^{\prime}$ that approximates the set of consistent worlds represented by $U$.
Queries over $U$ can instead be evaluated over $U^{\prime}$ achieving higher quality (due to reduced uncertainty in $U^{\prime}$) 
without additional complexity in query processing due to ICs. We demonstrate the effectiveness and scalability of our approach
to large datasets with complex constraints. We also present
experimental results demonstrating the utility of incorporating integrity constraints in uncertain relations, in the context of an
information extraction application.
\end{abstract}

\section{Introduction}
\label{sec:intro}
Recent advances in probabilistic models for information extraction, document classification, and automated tagging has revived significant interest in probabilistic data management. Extraction techniques based on models such as conditional random fields (CRFs) \cite{sunita} create a database wherein tuples and/or attribute values have associated explicit estimates of probability. Multiple probabilistic models \cite{tutorial,trio, abiteboul}, of varying expressivity, have been developed to represent such uncertain data along with efficient query processing approaches \cite{kochicde08, trio, tutorial} to support search and analysis capability on such uncertain databases. 
In this paper we develop an approach to incorporating additional semantics, in the form of database {\em integrity constraints}, that can reduce uncertainty in data, which, in turn, could positively impact applications such as query answering and retrieval 

Consider the example in Fig \ref{fig1} where we have an {\tt Employee} relation with uncertainty, represented using  "or-sets" \cite{tutorial} for each attribute .  For instance in tuple 1 the {\tt job-title} of the employee
"jim" is {\em either} {\tt instructor} (with probability $0.7$) or a {\tt manager} (with probability $0.3$). This uncertain relation represents 4 {\em possible worlds} which are the 4 possibilities of this relation based on different attribute value choices in each attribute. A query $Q$ over such a probabilistic database returns tuples that satisfy $Q$ in one of the possible worlds along with their corresponding probabilities. Now consider additional semantics in the form of say a functional dependency (FD) that states that a person cannot hold the same job title in 2 different divisions, i.e.,
{\tt (name, job-title)} $\rightarrow$ {\tt division}. Given this knowledge, we know that two out of the four possible worlds, where the first tuple has "Jim" as a "manager" (in "training"), are impossible as they violate the functional dependency. A natural extension to the query semantics is to return {\em only} those tuples that satisfy the query $Q$ in one of the {\em consistent} possible worlds. As a result, the answer to the query about "jim" above should not include the tuple 
$\{jim, manager, training\}$ since such a tuple is not part of any consistent instance of the relation. 

\begin{figure}[t]
\begin{tabular}{|l|l|l|l|} \hline
\multicolumn{4}{|c|}{\bf Employee} \\ \hline
{\bf name} & {\bf job-title} & {\bf division} & {\bf degree} \\ \hline
 jim & instructor(0.7) & training & MBA \\
  &  manager(0.3) & &  \\ \hline
 jim & manager & marketing & MBA \\ \hline
jim (0.5)& consultant & innovation & PhD \\
jill (0.5)  & & & \\ \hline
\end{tabular}
\\ 
{\bf Constraint (C):} \texttt{(name, job-title) $\rightarrow$ division}
\caption{Uncertain Relation with Constraints} 
\label{fig1}
\end{figure}

Incorporating additional knowledge such as constraints to reduce uncertainty in query results has indeed been explored to various degrees in different
uncertain database models and systems \cite{trio,chang07,koch08}. 
 For instance Trio \cite{trio} and \cite{chang07} permit the
specification of constraints, but at the data {\em instance} level i.e., between individual
attribute or tuple instances. For instance using the notation T1(2) to represent the second tuple instance (possibility) in the
first tuple in Fig \ref{fig1} i.e., {\tt (jim, manager, training, MBA)} and T2(1) to represent the first (and only)
tuple instance in the second tuple, we could specify a constraint such as T1(2) XOR T2(1) which states that only one
of these tuple instances can exist together in a possible world. Query answering approaches for such models
 \cite{chang07} can address only a small  number ($<$ 20) of such constraint instances. \cite{dalvis05} considers very restricted forms of FD and IND constraints in addition to database statistics, to address a different problem \-- that of determining a maximum likelihood {\em estimate} of the probability of a query answer in a data integration setting. 
MayBMS \cite{koch08} has considered more general integrity constraints at the level of individual tuples as well
as functional dependencies in their probabilistic model
based on representing uncertain databases using world set decompositions.
Their approach for factoring FDs however can be shown to be exponential - as we illustrate in the related work section.
This is not surprising as it can be shown that answering even simple selection queries exactly over uncertain relation 
in presence of integrity constraints (e.g., a single functional dependency), is NP-hard. We state the following:\\
{\bf Statement 1:} Given a U-relation, U, and a functional dependency (FD) F defined over U, identifying a possible world instance $pw_{q} \in PW_{U}$ such that $pw_{q} \models F$  or determining that no such instance exists is NP-Hard.
We refer to this problem as the {\bf The FD consistency problem.}\\
{\bf Proof:} This follows by a reduction from the 3-SAT problem which is known to be NP-Hard. The proof is as follows:\\

1. Given an instance of 3-SAT i.e., a CNF expression:\\
$(x_{11} \vee x_{12} \vee x_{13})$ $\wedge$ $(x_{21} \vee x_{22} \vee x_{23})$ ...$\wedge$ $(x_{n1} \vee x_{n2} \vee x_{n3})$. 
\\
We will now create a corresponding uncertain relation U as follows.\\
2. Consider any one conjunct, each conjunct is of the form of one of $(x_{1} \vee x_{2} \vee x_{3})$, $(x_{11} \vee x_{2} \vee \neg x_{3})$, $(\neg x_{11} \vee \neg x_{2} \vee x_{3})$, or $(\neg x_{11} \vee \neg x_{2} \vee \neg x_{3})$ \\
3. Take the following actions based on the type of the conjunct: \\ 
(i) Type is: $(x_{1} \vee x_{2} \vee x_{3})$ \\
Create the following 3 uncertain tuples, each tuple with 3 tuple instances (choices), in U: \\
\{$T_{x1}, T_{x2}, T_{x3}$\}\\ 
\{$T_{x1}, T_{x2}, T_{x3}$\}\\ 
\{$T_{x1}, T_{x2}, T_{x3}$\} \\
Let the tuple have some attributes (which are at least 3 in number). Let the first attribute be a tuple instance identifier (ID), tuple instance $T_{x1}$ has ID 1, tuple instance $T_{x2}$ has ID 2, etc. 
\\
(ii) Type is: $(x_{1} \vee x_{2} \vee \neg x_{3})$ \\
Note that we can also treat this as $x_{3}$ $\rightarrow $ ($x_{1} \vee x_{2}$)\\
Create the following tuple instances in U: \\
\{$T_{x1}, T_{x2}, $-$T_{x3}$\}\\ 
\{$T_{x1}, T_{x2}, $-$T_{x3}$\}\\
-$T_{xi}$ is a tuple instance created such that $T_{xi}$ and -$T_{xi}$ violate a functional dependency (FD). Pick 2 attributes A and B in the tuple. To inject an FD violation assign $T_{xi}$ as [i,..,a1,..,b1..] and -$T_{xi}$ as [i,..,a1,..,b2..] where a1 is a value of the A attribute and b1 and b2 are values of the B attribute. The instances $T_{xi}$ and -$T_{xi}$ thus violate the FD: A $\rightarrow$ B. 
\\
(iii) Type is: $(\neg x_{1} \vee \neg x_{2} \vee x_{3})$ \\ 
We can also treat this as $(x_{1} \wedge x_{2})$ $\rightarrow $ $x_{3}$
Create the following tuple instances in U: \\ 
\{-$T_{x1}, $-$T_{x2}, T_{x3}$\} 
\\ 
(iv) Type is: $(\neg x_{1} \neg \vee x_{2} \neg \vee x_{3})$ \\
We can also treat this as $(x_{1} \wedge x_{2})$ $\rightarrow $ $\neg x_{3}$
\\
Create the following tuple instances in U: \\ 
\{-$T_{x1}, $-$T_{x2}, T_{y}$\}\\ 
$T_{y}$ is made such that $T_{y}$ and $T_{x3}$ violate the FD: A $\rightarrow$ B. 
\\ 
At this point we have an instance of an FD consistency problem i.e., we have an uncertain relation U with uncertain tuples and a single FD: A $\rightarrow$ B on this relation. This reduction, from the original 3-SAT problem has been done in time polynomial in the size of the original 3-SAT problem.

Our claim is that a solution to the 3-SAT problem exists iff there is a solution to the FD consistency problem we have derived. Say we have a solution to the FD consistency problem. Each tuple is some $T_{xi}$. For each i, we can have only one of $T_{xi}$ or -$T_{xi}$ in the consistent relation obtained (so as to not violate the FD). For any $T_{xi}$ in the solution we set xi to 1 in the 3-SAT problem, for any -$T_{xi}$ in the solution we set xi to 0. With this assignment we will necessarily have a truth assignment for the xi s for which the 3-SAT formula is true. Also if there is a truth assignment that makes the 3-SAT formual true then there necessarily exists a solution to the FD consistency problem (for each xi assigned to 1 we retain $T_{xi}$ in the solution and for each xi assigned to 0 we retain -$T_{xi}$). Conversely if there is no solution to the FD consistency problem then there is no solution to the 3-SAT problem.

Our claim above is thus valid that a solution to the 3-SAT problem exists iff there is a solution to the translated FD consistency problem. As 3-SAT is NP-complete it follows that the FD-consistency problem is NP-Hard. \\

Given the intractibility of answering queries exactly in presence of ICs, we take a different approach that attempts to replace a given uncertain relation $U$ by another {\em sub-relation} that is also (a special case of) an
uncertain relation $U^\prime$  into which any constraints provided over $U$ have been {\em factored in}. Ideally, $U^\prime$ represents all the possible worlds of $U$ that are consistent w.r.t. $C$ and eliminates possible worlds that are inconsistent. The uncertain relation shown in Fig \ref{fig2} is such a $U^\prime$ for the relation $U$ in Fig\ref{fig1}. Answers to queries over $U^{\prime}$ ,which can be efficiently evaluated using independence semantics, would thus be exactly the answers were we to execute the query over consistent possible worlds of $U$. In general, such a $U^\prime$ that {\em exactly} captures the set of consistent possible worlds of $U$ might not exist. For instance if we modify the second tuple in the relation in Fig \ref{fig1} to be
\\
\begin{center}
\begin{tabular}{|l|l|l|l|} \hline
 jim & manager & marketing (0.5) & MBA \\ 
& & training(0.5) & \\ \hline
\end{tabular}
\\
\end{center}
we can see that {\em no} sub-relation (in the or-set based model we use) can exactly represent the consistent possible worlds of $U$. 
Our goal, thus, is to identify a ``good" sub-relation that
mirrors/approximates the original uncertain relation (and constraints) closely. Such a "good" approximation would eliminate as many of the inconsistent worlds of the original relation as possible while at the same time minimizing the number of consistent worlds that would invariably be eliminated as a by product. The paper devises mechanisms to computing such a "good" approximation for the original uncertain relation given a set of integrity constraints (IC). We consider a large class of attribute, tuple, and relation level ICs -  including
FDs, aggregation constraints and other kinds of ICs that other approaches have not considered.

Note that queries over a sub relation into which constraints have been factored can be answered efficiently. While simple selection queries over a single relation can be answered efficiently in a straightforward mechanism, techniques developed in \cite{dalvi07} can be used to answer more complex single as well as multi relation queries. Our specific contributions can be summarized as: (i) We present a more general approach for factoring a large class of ICs into uncertain databases that 
other systems have not considered, (ii) By using approximations our approach can scalably handle uncertain databases
with a high degree of data "dirtiness" (fraction of fields that are uncertain).

\begin{figure}
\begin{tabular}{|l|l|l|l|}
\hline
{\bf name} & {\bf job-title} & {\bf division} & {\bf degree} \\ \hline
jim & instructor (0.7) & training & MBA \\
\hline
jim & manager & marketing & MBA \\
\hline
jill (0.5) & consultant & innovation & PhD \\
jim (0.5) & & & \\
\hline
\end{tabular}
\caption{Alternate Representation}
\label{fig2}
\end{figure}

The rest of the paper is organized as follows. In Section~2, we formally define 
our notion of uncertain relations, 
state our problem of generating (tractable) sub-relations of uncertain
relations as part of our approach to providing efficient retrieval over uncertain relations with
constraints.  Section 3 and 4 together develops our approach where we borrow from and
build upon techniques from areas such as database repair \cite{bohannon} and work in compact
representation of probabilistic distributions \cite{sunita}. In Section 5, we demonstrate both the
scalability and efficiency of our approach as well as impact of considering ICs on quality of the
information extraction task. Section 6 gives an overview of related works and the last section
concludes the paper.

\section {Formalization}
In this section we formally define uncertain relations with constraints and postulate the problem of generating approximations of such uncertain relations that facilitate efficient query answering.  
\\
{\bf Uncertain Relation} An {\it uncertain relation}, $U$, is defined as:
\begin{itemize}
	\item $ U $ = $ \{ t_{1},t_{2},\cdots,t_{n}   \} $; i.e., $U$ is a relation which is a set of $n$ tuples.
	\item $t_{i} =  ( a_{i1},a_{i2},\cdots,a_{is}  ) $ ; each tuple is a sequence of $s$ attributes.
	\item $a_{ij} = \{(a_{ij}^{1},c_{ij}^{1}),...,(a_{ij}^{k_{ij}},c_{ij}^{k_{ij}} )\} $ ; Each attribute is a {\em set} of possible attribute values with an associated probability distribution. The set is referred to as the attribute world. $k_{ij}$ is the number of choices in the attribute world $a_{ij}$, and  $\sum_{p = 1}^{k_{ij}}  c_{ij}^{p} = 1 $ . 
\end{itemize}

Each uncertain relation $U$ represents a set of {\em possible worlds}, $PW_{U}$. Each possible world corresponds to choosing a value for each attribute $a_{ij}$, a specific value from its attribute world. Let $pw$ be a variable over the possible worlds. A possible world $pw$ = $pw_{q}$ $\in$ $PW_{U}$ is defined by a function, $f_{q}$ : $f_{q}(x,y) \rightarrow I $; where  $x \in \{1,2,..,n\}$, $y \in \{1,2,..,s\}$ and $I \in \{1,2, \cdots ,k_{xy} \}$. The number of such unique functions is $\prod_{i=1}^{n} \prod_{j=1}^{s} k_{ij} $ which is the number of possible worlds. The probability distribution $P_{I}$ defined over $PW_{U}$ under the assumption of independence is : 
\begin{equation}
\begin{split}
\forall pw_{q} \in PW_{U}, P_{I}(pw=pw_{q}) = \prod _{i=1}^{n} \prod_{j=1}^{s} c_{ij}^{f_{q}(i,j)}  
\end{split}
\end{equation} 
Note that $\sum_{all\, worlds\, q}^{} P_{I}(pw=pw_{q}) = 1 $. \\
The above model for representing database uncertainty is based on the \textit{or-set} relations \cite{imielinski} where an attribute value is essentially a set of possible values with an associated probability distribution. 
\\ \\
{\bf Uncertain Relation with Constraints} We now associate constraints with  uncertain relations, defining an uncertain relation {\em with constraints} denoted as $U+C$, where $U $ is an uncertain relation as defined above and $C$ is a set of integrity constraints over $U$. Let $PW_{U}^{C}$  denote the subset of possible worlds in $PW_{U}$ that are {\em consistent} w.r.t ({\em all}) the constraints, C. i.e., $PW_{U}^{C} = \{ pw_{q} | pw_{q} \in PW_{U}$ and $ pw_{q} \models C \} $. The set of possible worlds not consistent w.r.t. C is denoted as $PW_{U}^{\neg C} = \{ pw_{q} | pw_{q} \in PW_{U}$ and $ pw_{q} \not \models C \} $.
The uncertain relation with constraints, $U+C$, is interpreted as a set of possible worlds of $U$ with the probability distribution redefined as follows:

\begin{equation}\label{Di}
\begin{split}
P(pw = pw_{q}) & = 0,  if pw_{q} \not\models C \\
P(pw = pw_{q}) & = P(pw = pw_{q} | pw \in PW_{U}^{C} ), if pw_{q} \models C
       \\
 & = \frac{P(pw \in PW_{U}^{C}  \mid pw_{q}) P_{I}(pw = pw_{q})}{P(pw \in PW_{U}^{C})} \\
& (Bayes^{\prime} theorem) \\
& = \gamma P_{I} (pw = pw_{q}) 
\end{split}
\end{equation}
As $ P(pw \in PW_{U}^{C}  \mid pw_{q}) P(pw_{q}) =1$ since $pw \models C.$ 
Also $\gamma$ = 1/(1 - $\lambda$) where $\lambda = \Sigma_{pw_q} P_{I}(pw_{q}), pw_q \in PW_{U}^{\neg C}  $.
\\ \\
{\bf Sub-relations} Consider an uncertain relation $U$. If we replace the possible values of each attribute $a_{ij}$ in each tuple $t_{i}$ in $U$ with a {\em subset} of the possible values for that attribute in $U$, we arrive at what we call a {\em sub-relation} of $U$. We denote the sub-relation of $U$ by $U^\prime$. Strictly speaking $U^\prime$ is not an uncertain relation as it does not necessarily provide a complete probabilistic distribution over possible relations. It is used however to represent a subset of the possible worlds for an uncertain relation. A sub-relation $U^\prime$ is additionally associated with a constant factor $\gamma_{U^\prime}$ and the probability of any world $pw = pw_{q} \in PW_{U^\prime}$ is given by $p(pw=pw_{q}) = \gamma_{U^\prime} \prod _{i=1}^{n} \prod_{j=1}^{s} cij^{f_{q}(i,j)}$ i.e., the probability of any world is recalibrated with the  $\gamma_{U^\prime}$ factor.
The factor $\gamma_{U^\prime}$ is derived from Equation~\ref{Di} which ensures that the probability of any {\em consistent} world in $U^\prime$ is exactly the same as in $U+C$. Note however that $U^\prime$ may represent some inconsistent worlds as well and assign a non-zero probability to such worlds. 

As an example, Fig \ref{fig4} represents a sub-relation of the uncertain relation in Fig \ref{fig1} (and with the second tuple modified). $\lambda$ for the uncertain relation is 0.15. Thus $\gamma_{U^\prime}$ = 1/0.85 = 1.17 which is how the $\gamma_{U^\prime}$ factor for the sub-relation in Fig \ref{fig4} is derived. We define a sub-tuple of an uncertain tuple (any tuple in an uncertain relation is an uncertain tuple) analogously, where replacing the set of attribute values in each attribute in the tuple with one of its subsets provides us with a sub-tuple of that uncertain tuple.


We use sub-relations to approximate an uncertain relation $U$ with constraints $C$. Ideally, we would like the sub-relation $U^\prime$ to represent the exact set of consistent possible worlds of $U$ and to eliminate all of the inconsistent possible worlds. However, as discussed in the introduction, such a $U^\prime$ might not exist and, as a result, our goal will be to identify the "best" approximation of $U+C$. In order to define a concept of "best" we need to define a metric to evaluate how well does a  specific sub-relation capture $U+
C$. 

\begin{figure} 
\begin{tabular}{|l|l|l|l|} \hline
{\bf name} & {\bf job-title} & {\bf division} & {\bf degree} \\ \hline
 jim & instructor(0.7) & training & BA \\
\hline
 jim & manager & marketing (0.5) & MBA \\ 
& & training(0.5) & \\ \hline
jim (0.5)& consultant & innovation & PhD \\
jill (0.5)  & & & \\ \hline
\end{tabular} \\
$\gamma$ = 1.17
\caption{Sub-relation}
\label{fig4}
\end{figure}

{\bf Quality of Approximation:}. Let $U$ be an uncertain relation with associated integrity constraints $C$ and let $U^\prime$ be a sub-relation approximation of $U$. Let $P_c$ be the (total) consistent mass in $U+C$ (i.e., the sum of the probabilities of the possible worlds of $U$ that are consistent). Also, let $C_r$ ($I_r$) be the consistent (inconsistent) mass  of $U$ retained in $U^\prime$  respectively. The quality of $U^\prime$, denoted by $Q_{U^\prime}$ is defined as:
\begin{center}
$Q_{U^\prime} = \frac{C_r}{P_c} - I_r$
\end{center}
Note that this metric considers the {\em absolute} inconsistent mass retained and the {\em relative} consistent mass retained because it is the fraction of consistent mass retained that we would like to be high (as opposed to its absolute value which may be low). A quality value of 1 is the best achievable. 
We also note that since the approximate representation $U^\prime$ might eliminate consistent possible worlds (in addition to eliminating inconsistent worlds), the results of a query $Q$ over $U^\prime$ might include  false negatives (i.e., tuples that should be part of the answer since they satisfy the query in some consistent world, but are not part of the result over $U^\prime$). 
While introducing false negatives might be unacceptable for certain applications, for applications of probabilistic databases 
that motivate our work such as information extraction and query answering, we believe that a modest reduction in one of precision 
or recall in exchange for a significant increase in the other is a desirable tradeoff. 

{\bf Problem Formalization}
Given the above definition of quality, we can now formally state our objective as that of generating a sub-relation $U^\prime$ of $U$ that has the highest quality. That is, $\forall Y \in U^\prime_{M}$: $Q_{U^\prime} \geq Q_{Y}$, 
where $U^\prime_{M}$ is the set of "all" sub-relations. Unfortunately, identifying such an "optimal" sub-relation is NP-hard even when we consider a single functional dependency or a tuple level constraint as we will see in the next section ~\cite{techrep}. We will therefore restrict ourselves to heuristic techniques to finding $U^\prime$ that attempt to maximize $Q_{U^\prime}$.

\section{Incorporating ICs in an Uncertain Relation}

In this section, we describe our approach to generating the approximate sub-relation $U^\prime$
given an uncertain relation $U$ and a set of constraints $C$ that hold over $U$. Our approach
starts with the original relation $U$, selects a constraint $C_i \in C$, and attempts to resolve $C_i$ 
by dropping (a minimal number) of attribute values from tuples in $U$ such that the resulting
sub-relation does not violate $C_i$. The process of resolving constraints (or "fixing" the relation $U$) is iteratively carried out until the
algorithm deems that the benefit of further removing inconsistency no longer outweigh the loss
of the consistent worlds that results as a by-product of "fixing" the uncertain relation. Before we
discuss the details of the algorithm, we first need to specify the nature of integrity constraints
(IC) that we consider in the paper. The approach we use to fix the uncertain
relation depends upon the nature of the  integrity constraint. 

We classify ICs into the following three different categories: \\
{\bf (i) Attribute level ICs:} Constraints that depend on the
    values of a specific attribute in a tuple, and not on other attributes 
    in the same tuple or other tuples. An
    example can be 
{\tt CHECK degreelevel(degree)} 
 that states, through a user defined function (UDF), that the value for the {\tt
degree} must be at least a 4-year college degree. We will assume that attribute level ICs can be
checked efficiently (in polynomial time). \\
{\bf (ii) Tuple level ICs:} Constraints that are dependent on the values
    of two or more attributes within a specific tuple, and not on the
    values of attributes of different tuples. As an example: 
{\tt CHECK compatible(job-title,degree)} 
 may represent a tuple level IC that enforces, also through a UDF, some
compatibility between  a person's job title and his degree (e.g., that a "manager"
must have at least an "MBA" degree, etc.). We will assume that each
instance of a tuple-level IC can be checked for constraint violation efficiently
(in polynomial time.) In addition, we will assume that the arity of the constraint, i.e. number of
attributes associated with the constraint is small enough such that enumerating all tuple instances
that could be potential constraint violations is tractable.
\\
{\bf (iii) Relation level ICs:} Constraints that exist across different
    attributes from different tuples. 
For instance a constraint:  \\ 
{\tt CREATE ASSERTION no-multiple-divisions \\ CHECK (SELECT COUNT division FROM employees GROUP BY (name, job-title) == 1)} \\
states that the same person cannot have the same job-title in two different
divisions. This constraint is essentially the FD {\tt (name, job-title)
$\rightarrow$ (division)}. Note that {\tt "Check"} constraints at the attribute
level (or at the tuple level) that depend upon other tuples will also be classified
as relation level constraints.

The set of constraints, $C$, is a union of attribute level, tuple level, and
relational level constraints, represented as $C_a$, $C_t$, and $C_r$ respectively  i.e.,  
$C = Ca \cup Ct \cup Cr$. 

We next discuss our strategies to resolve attribute, tuple, and relation level constraints
independently. After describing our strategies to resolve single constraints, we will describe our
algorithm to resolve the set of constraints  $C$. In the remainder of the section, we will use the
example uncertain relation with constraints in Figure \ref{fig5} as an example for illustration.
 
\begin{figure}[t]
\vspace*{0.05cm} 
{\bf Relation: U } \\ \\
\begin{tabular}{|l|l|l|l|}
\hline
{\bf name} & {\bf job-title} & {\bf division} & {\bf deg} \\
\hline
jim & instructor (0.7) & training & BA (0.2) \\
& manager (0.3) & & MBA (0.8) \\ 
\hline
jim & manager & marketing & MBA \\
\hline
jill (0.5) & consultant & innovation & AAB (0.4) \\
jim (0.5) & & & PhD (0.6) \\
\hline 
\end{tabular}
\vspace*{0.3cm}
\\
{\bf Constraints: C} \\ \\
Attribute level ICs (Ca) \\
1. {\tt CHECK degreelevel(deg)} \\
{\em All employees have at least a 4 year college degree.} \\ 
Tuple level ICs (Ct) \\
1. {\tt CHECK compatible(division,deg)} \\
{\em All "training" division employees have at least an "MBA" degree.} \\ 
Relation level ICs (Cr) \\
1. {\tt CHECK (name, job-title) $\rightarrow$ division} \\
{\em An employee does not hold the same title in 2 different divisions} 
\\
\\
\caption{Uncertain Relation with Constraints}
\label{fig5}
\end{figure}

\begin{center}
\begin{table}[t]
\begin{tabular}{|l|l|l|l|}
\hline
jim & instructor (0.7) & training & BA (0.2) \\
& manager (0.3) & & MBA (0.8) \\ 
\hline
jim & manager & marketing & MBA \\
\hline
jill (0.5) & consultant & innovation & PhD (0.6) \\
jim (0.5) & & & \\
\hline
\end{tabular}
\\
\vspace*{0.2cm}
\\
(a) $U_{1}$: Factored attribute levels ICs
\\ \\
\begin{tabular}{|l|l|l|l|}
\hline
jim & instructor (0.7) & training & MBA (0.8) \\
& manager (0.3) & &  \\ 
\hline
jim & manager & marketing & MBA \\
\hline
jill (0.5) & consultant & innovation & PhD (0.6) \\
jim (0.5) & & & \\
\hline
\end{tabular}
\\
\vspace*{0.05cm}
\\
(b) $U_{2}$: Factored tuple level ICs
\\ \\
\begin{tabular}{|l|l|l|l|}
\hline
jim & instructor (0.7) & training & MBA (0.8) \\
& & &  \\ 
\hline
jim & manager & marketing & MBA \\
\hline
jill (0.5) & consultant & innovation & PhD (0.6) \\
jim (0.5) & & & \\
\hline
\end{tabular}
\\
\vspace*{0.05cm}
\\
(c) $U^{\prime}$: Factored relation level ICs, final approximation
\\
\caption{Generating Approximations}
\label{tab1}
\end{table}
\end{center}
Resolving attribute level ICs is actually tivial as in any attribute world we simply eliminate
any attribute instance that is not consistent with an IC in $C_a$. This is illustrated in 
table \ref{tab1} (a) where the \textbf{AAB} value in tuple 3 is dropped. We note that the sub-relation that results from 
resolving $C_a$ removes only the inconsistent worlds but does not remove any consistent ones. 

\subsection{Resolving A Tuple Level IC}

To resolve a tuple level constraint $C_{tup} \in C$, we can consider each tuple $T$  of the
uncertain relation independently. Given an uncertain tuple $T$ and a specific tuple level
constraint $C_{tup}$, we would ideally like to arrive at a sub-tuple $T^{\prime}$ (of $T$) that
is equivalent to $T+C_{tup}$, i.e. it satisfies $C_{tup}$, while allowing the same set of possible
consistent instances as $T$.
Unlike the case of attribute level IC, dropping attribute values from tuples in $U$ that
violate $C_{tup}$ might result in one or more consistent instances to be eliminated as well. As a
result, the resulting sub-relation $U'$ might not exactly represent the set of consistent possible
instances in $U+C_{tup}$. Fig \ref{utup} illustrates such an example with a constraint that all
training division employees have at least an "MBA" level degree. Dropping any attribute value
from the tuple results in a loss of a consistent instance. For instance, removing "BA" from the
attribute world of {\tt degree} attribute results in a sub-tuple that satisfies the considered tuple
level IC, but it eliminates the consistent possible instance in which "jim" works in "marketing"
division with a "BA" degree. Furthermore, the problem of identifying the sub-relation that
optimally approximates  (in terms of quality) the set of possible worlds of the uncertain
relation $U$ consistent w.r.t. a single tuple level constraint $C_{tup}$ remains NP-hard. We state the 
following:\\
{\bf Statement 2:} {\em Determining an optimal approximation T$\prime$ of an uncertain tuple T is NP-Hard} \\
{\bf Proof:} This follows by a reduction from the functional dependency (FD) consistency problem. The proof is as follows:\\
1) Consider any given instance of an FD consistency problem (U,F) where U is a U-relation and F is an FD over U. \\
2) Create a new tuple, T, as follows. For every tuple $t_{i}\in$ U create a new attribute $A_{t_{i}}$ in T. For each tuple {\em instance} $t_{i}^{k}$ in every tuple $t_{i}$ in  U, create a corresponding attribute value instance in $A_{t_{i}}$. Finally, provide a uniform probability distribution in all attribute worlds in T. \\
3) For every instance of a pair of tuple instances $t_{i}^{m}$ and $t_{i}^{n}$ (i$\neq$j) that violate F, create an instance of a constraint violation between the corresponding attribute value instances in T. \\
4) T is an uncertain tuple that possibly also has some constraint violations across attribute values. Note that the reduction from the FD consistency problem to this uncertain tuple T is done in time polynomial in the size of the original problem. \\
5) Generate an optimal approximation T$\prime$ of T. If there is {\em any} tuple instance that is consistent in T then at least one such consistent instance {\em must} appear in T$\prime$. This is because all consistent tuple instances in T have the same probability and all inconsistent instances have a probability of 0. Also if T$\prime$ is empty then this implies that there are no consistent tuple instances whatsoever in T. \\
6) The tuple instances in T directly correspond to relation instances in the original FD consistency problem as there is a 1-1 mapping from the attribute values instances in attributes in T to tuple instances in tuples in U. Any consistent tuple instance in T directly corresponds to a consistent {\em relation instance} in U. \\
7) The original problem of determining a consistent relation instance in U (or determining that none exists) is however NP-Hard. This implies that the problem of optimal tuple approximation, that this was reduced, to is also NP-Hard. \\

Given the intractability of identifying the optimal sub-relation, we focus on developing a
heuristic approach to find a "good" approximation that preserves as much of consistent mass as
possible (see Sec. 2) which we describe next. 

\begin{figure}[t] 
\begin{tabular}{|l|l|l|l|} 
\hline
{\bf name} & {\bf job-title} & {\bf division} & {\bf degree} \\ \hline
 jim & instructor & training (0.6) & BA (0.7) \\
 & & marketing (0.4) & MBA (0.3) \\
\hline
\end{tabular} 
\caption{Sample U-tuple, for which no proper sub-tuple exists}
\label{utup}
\end{figure}

\begin{figure}[t]
\centering
\includegraphics[width=3.0in]{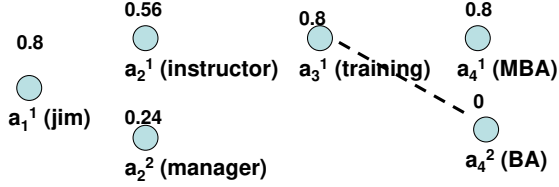}
\caption{Graph Representation of Uncertain Tuple}
\label{fig13}
\end{figure}

\begin{tabular}{p{7.5cm}}
\\ \hline
\end{tabular}
\begin{tabbing}
{\bf Algorithm: APPLY\_TUPLE\_IC} \\
{\em Input:} Uncertain relation $U_0$, Tuple level IC $C_{tup}$ \\
{\em Output:} Sub-relation $U_1$ \\
1: APPLY\_TUPLE\_IC\_SR ($U_0,C_{tup}$) \\
2: $t_{new}$ $\leftarrow$ $\o$ \\
3: {\bf for} \= (each tuple t in $U_0$) \\
4: \> $t_{new}$ $\leftarrow$ $t_{new}$ $\cup$ APPLY\_TUPLE\_IC\_SR\_TUPLE(t,
$C_{tup}$) \\
5: $U_{1}$ $\leftarrow$ form\_relation($t_{new}$) \\
6: {\bf return} $U_{1}$ \\
\\
1: APPLY\_TUPLE\_IC\_SR\_TUPLE (T, $C_{tup}$) \\
2: \>          ATTRIBUTE\_MARGINALS(T,S) \\
3: \>          G $ \leftarrow $  graph\_representation(T, $C_{tup}$)  \\
4: \>    I $\leftarrow $ independent\_nodes(G) \\
5: \>    N $\leftarrow $ $G  - I$ \\
6: \>    Nb $\leftarrow $ best\_candidate\_to\_delete(N) \\
7: \>    G $\leftarrow$ delete(G,Nb) \\
8: T $ \leftarrow $  tuple\_representation(G,$\gamma$)  \\
9: return(T) \\
{\bf form\_relation:} Construct a new relation. \\
{\bf graph\_representation:} Convert uncertain tuple to graph.\\
{\bf independent\_nodes:} Find nodes without any edge. \\
{\bf best\_candidate\_to\_delete:} Find proper node to remove.\\
{\bf tuple\_representation:} Convert graph to tuple. 
\end{tabbing}
\begin{tabular}{p{7.5cm}}
\\ \hline
\end{tabular}
\\
\\
For a given tuple $T$ of $U$ and a tuple level constraint $C_{tup}$, we start with
constructing a 
graph representation of $T$ in which nodes correspond to attribute value instances in
each attribute,  and edges and hyper-edges represent sets of attribute value instances (across
attributes) that violate the tuple level constraints. The graph representation of the first tuple in Fig \ref{fig5} is shown in Fig \ref{fig13}. We now delete nodes in this graph till all the (hyper)
edges disappear, the resulting graph represents the attribute value instances that are consistent
w.r.t. $C_{tup}$ and can hence be retained in the approximation. For choosing nodes to drop,
recall that we are interested in approximations with high quality i.e., where any consistent mass
dropped is minimal. The consistent mass associated with any individual node (attribute value) is
given by its {\em marginal} probability in the tuple. The marginal probability of an attribute
value instance $a_{ij}^{k}$, denoted as $p_{MARG}(a_{ij}^{k})$ is defined as the sum of the
probabilities of all the tuple instances implied by the uncertain tuple that include that attribute
world instance.

\begin{equation}
p_{MARG}(a_{ij}^{k}) = \sum_{all\, instances\, t \in T \, \wedge \, a_{ij}^{k} \in t} p(t) 
\end{equation}

We adorn the graph nodes with their associated marginal probabilities. We then choose the nodes
to drop in a greedy fashion biasing towards dropping nodes with low marginal probabilities, till
all (hyper) edges have been eliminated. As an example, consider again the graph in figure
\ref{fig13}, and its corresponding sub-tuple. Having just one tuple level IC, and a pair of
violating possible attribute values, we can eliminate the only existing inconsistency, shown as the
dashed edge in the graph, by dropping one of its corresponding nodes. In this case, we decide to
drop $a_{4}^{2}$, the "BA" value, according to its marginal probability, which is 0 comparing
to the marginal probability of $a_{3}^{1}$, "training" value, which is 0.8.The complete algorithm
is described in Algorithm APPLY \_TUPLE\_IC. 

Note that our approach requires that the  marginal probability value for each attribute value instance
in the tuple be known. Unfortunately, computing the marginal probabilities of attribute values
instances in an uncertain tuple can be shown to be NP-Hard. Instead, we can {\em
estimate} such marginals using statistical sampling. We employ naive-MC (Monte-Carlo)
sampling. The procedure for estimating marginal probabilities of attribute value instances in a
tuple, based on sampling randomly generated tuple instances,  is described in algorithm
ATTRIBUTE\_MARGINALS.
\\
\\
\begin{tabular}{p{7.5cm}}
\\ \\ \\ \hline
\end{tabular}
\begin{tabbing}
{\bf Algorithm: ATTRIBUTE\_MARGINALS} \\
{\em Input:} Uncertain tuple T, Number of samples S \\
1: ATTRIBUTE\_MARGINALS(T,S) \{ \\
2: {\bf for} \= (all attribute value instances $a_{ij}^{k}$ in all attributes in T) \\
3: \> $p_{MARG}(a_{ij}^{k}$) $\leftarrow$ 0 \\
5: {\bf for} \= (i = 1 through S) \\
6:     \> $t_{samp}$ $\leftarrow$ random\_sample(T) \\
7:     \> for \=(all attribute value instances $a_{ij}^{k}$ $\in$ $t_{samp}$) \\
8:     \> \> $p_{MARG}(a_{ij}^{k}$)  $\leftarrow $ $p_{MARG}(a_{ij}^{k}$)  +
$p(t_{samp})$ \\
10: for \=(all attribute value instances $a_{ij}^{k}$ $\in$ T) \\
11: \> $p_{MARG}(a_{ij}^{k})$ $\leftarrow$ $p_{MARG}(a_{ij}^{k}$)/S \\
12: return($\{p_{MARG}(a_{ij}^{k}\}$ ) \\
{\bf random\_sample(T):} random tuple instance. \\
\end{tabbing}\begin{tabular}{p{7.5cm}}
\\ \hline
\end{tabular}
\\
\\
{\bf Statement 3:} {\em The derivation of marginal probabilities of attribute value instances in an uncertain tuple, or of tuple instances in an uncertain relation with constraints, is NP-Hard}. \\
{\bf Proof:} Given an instance (U,F) of the FD consistency problem we determine the marginal probabilities of the tuple instances in each tuple in U. A consistent relation instance in U exists iff the marginal probability of at least one of the tuple instances (in any tuple in U) is $>$0. The original FD consistency problem is however NP-Hard. This implies that determining the marginal probabilities of tuple instances in tuples in a U-relation is also NP-Hard \\
\\
For determining the complexity of determining marginal probabilities of attribute values in an uncertain tuple we make a reduction from the FD consistency problem. Given an instance of the FD consistency problem (U,F) we create an uncertain tuple, T, exactly as in the proof for Statement 2 above. A consistent instance in the FD consistency problem is present iff the marginal probability of (at least) one of the attribute value instances in T is $>$ 0.  The original FD consistency problem is however NP-Hard. This implies that determining the marginal probabilities of attribute value instance in attributes in an uncertain tuple is also NP-Hard. \\

\subsection{Resolving A Relation Level IC} 

For a relation level IC the instances of violations of that IC could be exponential in the number of tuples.
The approach we used for resolving tuple level ICs - which involves exhaustively enumerating and imprinting
all instances of violations, is thus not practical for relation level ICs. Also in the context of a relation level IC, we will use the term "tuple instance" to refer to the projection of the tuple instances onto the attributes that are part of the IC. Resolving a specific relational level IC, $C_{rel}$, in an uncertain relation $U$, comprises the
following two steps: \\ 
a) Within $U$ we identify {\em sets} of tuple instances where each set can potentially violate
$C_{rel}$. For instance for a functional dependency IC, $A \rightarrow B $, where $A$ and $B$ are
two sets of attributes according to the schema of $U$, any set of tuple instances which agree on the
value of $A$, form a set of tuple instances that could potentially violate the FD. A
possible relation of $U$, where tuple instances are drawn from such a set, could be inconsistent with
$C_{rel}$. Given any $C_{rel}$, all such sets of tuple instances can be determined exhaustively (the number of such sets is proportional to the number of distinct attribute values of $A$).
We refer to any such a set as a "NEED-FIX" class for that $C_{rel}$. As an example, a
NEED-FIX class for the FD constraint over the uncertain relation in Fig \ref{example} (a) is illustrated in Fig \ref{example} (b). The tuple instances are denoted by first specifying the tuple number in the uncertain
relation and then the tuple instance number within each tuple in ().\\
b) We eliminate the inconsistencies in any NEED-FIX class considering each class individually.
This is achieved by {\em dropping} tuple instances in the class till consistency is achieved. We
refer to this as "fixing" a class. For instance for a NEED-FIX class corresponding to a functional
dependency $A \rightarrow B$, we would drop tuple instances until all the tuple instances in that
class agree on the value of the attribute(s) in $B$. Note that in general there may be many different
combinations of tuple instances that can be dropped that will achieve consistency. For instance, the NEED-FIX class
in Fig \ref{example} (b) can be fixed by dropping either the 1st tuple instance, or the 2nd and 3rd tuple
instances in the class.

\begin{table*}[t]
\begin{minipage}[t]{0.9\textwidth}
\small \centering
\begin{tabular}{|p{3cm}|p{5cm}|p{5cm}|p{2cm}|}
\hline
{\bf CONSTRAINT TYPE} & {\bf Generating NEED-FIX class(NF)} & {\bf Fixing NF} & {\bf Complexity*} \\ 
\hline
{\em Type:} Functional Dependency (FD) &  1) For each tuple instance t in each tuple
T in U. & 1) Group the tuple instances by the value of B & $O(N_{t}) $\\
{\em Format:} {\tt A $ \rightarrow$ B} where A,B are subsets of columns in U &
2) Initialize a new NEED-FIX class, NF, with the single member t. \newline 3) For
any tuple $T\prime$ in U that contains \newline a tuple instance $t\prime$ such that
$t\prime$.A=t.A, \newline add $t\prime$ to NF. \newline 4) Add NF to the pool of
NEED-FIX \newline classes. & 2) Select the value for B for which the sum of the
marginals (of the tuple instances) in that group is the highest. \newline 3) Drop
all tuple instances with \newline values for B other than the above \newline
selected value. & \\
\hline
{\em Type:} Inclusion Dependency(IND) & 1) Initialize a new NEED-FIX class,NF, to
NULL. & 1) Drop all tuple instances in NF. & $O(N_{t})$  \\
{\em Format: } {\tt U.A $\in $ E.B} where E is a fixed relation and
A, B are subsets of columns in U and E respectively& 2) For any tuple instance t in
tuple T, if t.A $\neg \in $ E.B then add T to NF. \newline 3) Add NF to the pool of
NEED-FIX classes. & & \\
\hline
{\em Type:} Aggregation \newline {\em Format:} {\tt GROUP BY A COUNT $ <$ G  };where A is a subset of columns in U and G
is an integer.&  1) For
each tuple instance t in each tuple T in U. \newline 2) Initialize a new NEED-FIX
class,NF, with the single member t. \newline 3) For any tuple $T\prime$ in U that
contains a tuple instance $t\prime$ such that $t\prime$.A=t.A, add $t\prime$ to NF.
\newline 4) Add NF to the pool of NEED-FIX classes. & 1) Let Nnf be the number if
tuple instances in NF. \newline 2) If Nnf $<$ G then we are done. \newline 3) Else
Nnf $-$ G tuple instances have to be dropped. Drop those Nnf $-$ G tuple instances
from NF for which the sum of the marginal values is minimum. &
$O(^{N_{T}}C_{N_{T}\-G})$ \\
\hline
{\em Type:} Aggregation \newline {\em Format: } {\tt GROUP BY A EXP(B) $\theta$ val}; where {\tt EXP} is one
of \{AVERAGE, SUM, COUNT\}, A is a subset of columns in U, and B is a (numeric) column in U, and $\theta$ is one of $\{=,\leq, <\} $  & Same as above. & 1) Exhaustively search all
combinations of tuple instances that can be dropped to make NF consistent wrt this
constraint. \newline 2) Determine the combination with the minimum total marginal
value and drop the tuple instances in that combination. & $O((2^{P})^{N_{T}})$ \\
\hline
{\em Type:} SET Constraint \newline {\em Format:} {\tt Q $ \theta $ E.B} ; where {\tt  Q =
(SELECT A FROM U WHERE CND)}, CND is a query condition, and $\theta$ is one of $\{=,
\leq, <\} $ & 1) Initialize a new NEED-FIX class,NF, to NULL. \newline 2) For any
tuple instance t in tuple T in the result of Q, if t.A $\in $ E.B then add T to NF.
\newline 3) Add NF to the pool of NEED-FIX classes. & 1) Drop all tuple instances in
NF. & $O(N_{t}) $ \\
\hline
\end{tabular}
\label{tab2}
\end{minipage}
\centering 
\caption{Addressing Relation Level Constraints (\textsf{$N_{t}$: Total number of
tuple instances in NF; $N_{T}$: Total number of tuples represented in NF; $P$:
Maximum number of tuple instances in any tuple in NF.})}
\end{table*}

\begin{figure}[htp]
\begin{center}
\begin{tabular}{|l|l|l|}
\hline
{\bf name} & {\bf job-title} & {\bf division} \\
\hline
jim & instructor (0.5) & marketing \\
& consultant (0.5) &  \\ 
\hline
jim & instructor (0.3)& training \\
& manager (0.7) &  \\
\hline
jim & instructor & training  \\
\hline 
\end{tabular} \\
\end{center}
ICs: \\
1. \small{CHECK (name, job-title) $\rightarrow$ division} \\
2. \small{CHECK GROUP BY (name, job-title) COUNT * $<$ 2}\\
(a) Example uncertain relation with constraints \\ \\

\begin{tabular}{cc}
\begin{tabular}{|l|}
\hline
1(1) jim instructor marketing  \\
2(1) jim instructor training \\
3(1) jim instructor training \\ 
\hline
\end{tabular}

&

\begin{tabular}{|l|}
\hline
2(1) jim instructor training \\ 
3(1) jim instructor training \\
\hline
\end{tabular} \\
(b) NEED-FIX class: FD IC &
(c) NEED-FIX class: Aggregation IC  \\
\end{tabular}
\caption{An Example}
\label{example}
\end{figure}

The process of determining a NEED-FIX class, fixing it, and also the computational complexity
of the fix operation are dependent on the {\em type} of the relational constraint that is being
addressed. We described the generation and fixing of NEED-FIX classes for FD type ICs above.
For aggregation constraints, such as the 2nd IC in the example in Fig \ref{example} (a), NEED-FIX classes
are determined by grouping together tuple instances that agree on the attributes that we must
{\em group by} according to the aggregation constraint.  One such NEED-FIX class is shown in
Fig \ref{example} (c) where we have grouped together tuple instances by {\tt (name, job-title)}. The fix is a
process of eliminating tuple instances such that the aggregation constraint {\em condition} is
satisfied, in this example dropping either of the tuple instances in the NEED-FIX class will
ensure this. Table II presents the specific kinds of relation ICs addressed and the associated complexity.
The procedures for generating and fixing NEED-FIX classes for IND and SET constraints are straightforward and we do not present them here for lack of space.
As we have seen there can be multiple sets of tuple instances that can be dropped to fix a
NEED-FIX class. The choice of an optimal set of tuple instances to drop is made based on the
marginal probabilities of each tuple instance. Formally, the marginal probability,
$p_{MARG}(t)$, of a tuple instance, t in an uncertain relation U is defined as: 

$p_{MARG}(t)$ = $\Sigma_{all\, instances\, u \in U}$ $p(u)$ ; $ t \in u$.  Like attribute marginals, the derivation of tuple instance marginals in a relation is also NP-
Hard\cite{techrep}. We employ naive-MC sampling for estimating these marginals in a manner
analogous to the attribute marginals estimation, and here  we sample randomly generated
relation instances.

For any NEED-FIX class we can determine the combination of tuple instances with lowest (total) marginal probability, that if dropped will eliminate the inconsistencies in that class. The complexity of resolving a NEED-FIX class is polynomial in the size of the NEED-FIX class for (the permitted) FDs, INDs and SET constraints and is exponential (in the size of the NEED-FIX class) for the aggregation constraints. For aggregation constraints we use a simple hill-climbing procedure to find a set of tuple instances to drop that will remove the inconsistency in the NEED-FIX class and also have a low (total) marginal probability. 
\begin{tabular}{p{7.5cm}}
\\ \hline
\end{tabular}
\begin{tabbing}
{\bf Algorithm: APPLY\_RELATION\_IC} \\
{\em Input}: Uncertain relation $U_0$, Relation level IC $C_{rel}$ \\
{\em Output:} Sub-relation $U_1$ \\
1: APPLY\_RELATION\_IC($U_{0}, C_{rel}$)  \\
2: $NF$ $\leftarrow$ $\o$ \\
3: $\gamma$ $\leftarrow$ estimate\_gamma($U_{0}, C_{rel}$) \\
4: {\bf for} \= (each tuple instance t in each tuple T in $U_{0}$) \{ \\
5:   \>     $NF_t$ $ \leftarrow $ generate\_need\_fix\_class(t,$U_{0}$, $C_{rel}$) \\
6:   \>   NF $\leftarrow$ NF $\cup$  $NF_t$     \\
7: \} \\ 
8: $NF$ $\leftarrow$ FIX(NF) \\
9: $U_1$ $\leftarrow$ form\_relation($NF$, $\gamma$) \\
10: {\bf return} $U_1$ \\
{\bf generate\_need\_fix\_class:} generates a new NEED-FIX class \\given a tuple
instance and a relation level IC. \\
{\bf fix:} fix a particular NEED-FIX class \\
{\bf form\_relation:} form a new relation. \\
\end{tabbing} 
\begin{tabular}{p{7.5cm}}
\\ \hline
\end{tabular}
\\
\\

\section{Using a Multi-Row Representation}
Revisiting the example in Fig \ref{utup} we realized that in order to achieve consistency (by dropping some instances) some loss of consistent instances was invariable. This is because of the model simplicity and we have been using what is called a single-row model \cite{sunita}. A representation model that permits multiple rows for each tuple, known as a multi-row model, can overcome this limitation as illustrated in Fig \ref{mrex} where the approximation now exactly captures the uncertain tuple in Fig \ref{utup}.
\begin{figure}[htp]
\begin{tabular}{|l|l|l|l|}
\hline
jim & instructor (0.7) & training (0.6) & MBA (0.8) 1 \\
& manager (0.3) & marketing (0.4) & \\
\hline
jim & instructor (0.7) & & MBA (0.8) 1\\
& manager (0.3) & marketing (0.4) & BA (0.2)\\
\hline
\end{tabular}
\caption{Uncertain Tuple Approximation}
\label{mrex}
\end{figure}
\\
\begin{figure}
\centering
\includegraphics[width=\linewidth]{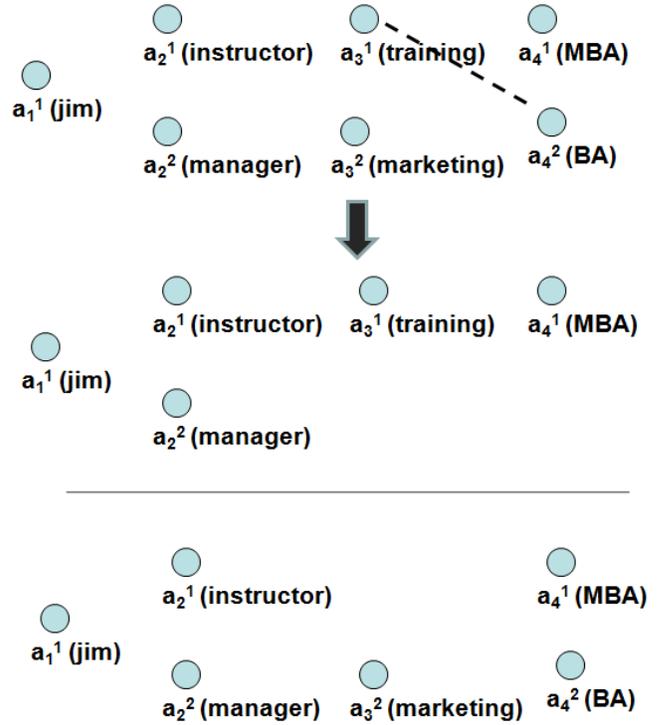}
\caption{Multi-Row Example}
\label{mr}
\end{figure}
While in the above example the multi-row representation exactly captured the uncertain tuple with a small number of rows (2), this is not the case in general. We present the following: \\ \newline
{\bf Statement:}  {\em The number of rows in a multi-row representation required to exactly capture an uncertain tuple with constraints can be exponential (in the size of the largest attribute world in the tuple) in the worst case.} \\
{\bf Proof:} Given a tuple and a set of constraint violations (let us consider only binary constraints violations across pairs of attribute values wlog) assume that there is a multi-row representation with $M$ rows. Consider any row, $r$, where we have at least one attribute that has at least 2 attribute values. We {\em insert} a new violation between any of these (multiple) attribute values and any attribute value in any other attribute in the tuple. Now $r$ must necessarily be split into at least 2 rows to exactly capture the consistent tuple instances. We can continue inserting violations in rows in this manner with an upper bound of $KC_{2}A^{2}$ violations that we will insert where K is the number of attributes. The number of rows that we will form in the multi-row model can however be as much as $O(A_{K})$ i.e., exponential in the (maximum) size of the attribute worlds in the tuple. \\
An approximation that takes exponential space is not tractable to reason with and we are interested in multi-row approximations where the number of rows is bounded by a constant or at least a factor that is polynomial in the size of the original uncertain tuple. With such a restriction we can at best achieve an optimal approximation as opposed to an exact one in the general case. Any multi-row approximation is defined by 2 kinds of {\em parameters}, one is the number of rows in the representation and the other is the assignment of probabilistic values to attribute value instances within each attribute within each row.  The complexity of deriving an optimal approximation is an issue however, we present the following:
\\
{\bf Statement:} {\em For a multi-row representation with a bounded number of rows, determining multi-row model parameters that result in an optimal approximation of a tuple is NP-Hard.} \\
{\bf Proof:} This too follows from a reduction from the FD consistency problem, and the proof is analogous to as for the single row model. \\

\begin{figure}
\centering
\includegraphics[width=\linewidth]{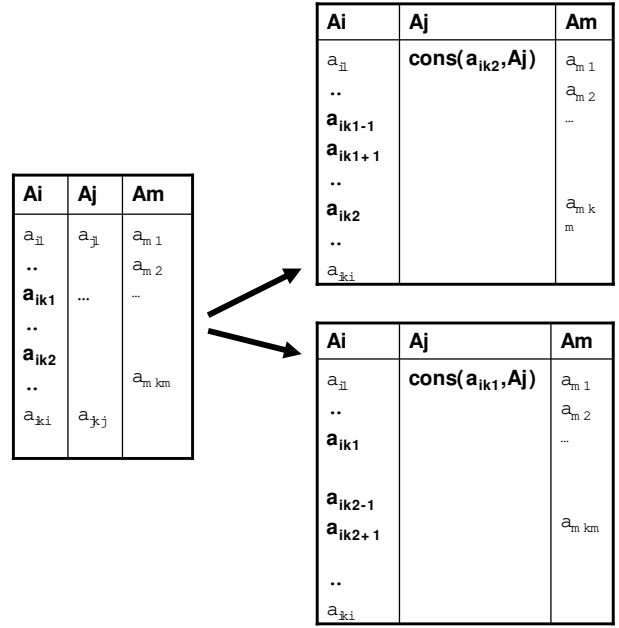}
\caption{Split to Multi-Row}
\label{fig6}
\end{figure}

\begin{tabular}{p{7.5cm}}
\\ \hline
\end{tabular}
\begin{tabbing}
{\bf Algorithm: APPLY\_TUPLE\_IC\_MR} \\
{\em Input:} Uncertain relation $U_0$, Tuple level IC $C_{tup}$ \\
{\em Output:} Sub-relation $U_1$ \\
1: APPLY\_TUPLE\_IC\_MR ($U_0,C_{tup}$) \\
2: $t_{new}$ $\leftarrow$ $\o$ \\
3: {\bf for} \= (each tuple t in $U_0$) \\
4: \> $t_{new}$ $\leftarrow$ $t_{new}$ $\cup$ APPLY\_TUPLE\_IC\_MR\_TUPLE(t,$C_{tup}$) \\
5: $U_{1}$ $\leftarrow$ form\_relation($t_{new}$) \\
6: {\bf return} $U_{1}$ \\
\\
1: APPLY\_TUPLE\_IC\_MR\_TUPLE(t,$C_{tup}$,M) \\
2: $V \leftarrow $ determine\_violation\_sets(T) \\
3: m=0, F=0
4: {\bf while} \= (m $< $M and inconsistent(T))  \\
6: \>    $ TopV \leftarrow top\_violation\_set(V) $ \\
7: \>    $ T \leftarrow split(T,TopV)$ \\
8: {\bf end while} \\
9: {\bf for}  \= (each row R $\in$ T) \\
10: \> $R \leftarrow $ APPLY\_TUPLE\_IC\_SR(R)  \\
11: {\bf end for} \\
12: return T \\
\end{tabbing}
\begin{tabular}{p{7.5cm}}
\\ \hline
\end{tabular}
\\
\\
What we employ is a heuristic approach to generating a multi-row approximation for a given uncertain tuple. We describe our approach using the example of binary tuple level constraints although the basic approach is valid for general (k-ary) tuple level constraints. Continuing with the graph representation of a tuple as described earlier, we recall that our aim
was to eliminate (hyper) edges in the graph by dropping nodes. In the multi-row model our aim is to instead 
{\em split} the graph into multiple sub-graphs such that the (hyper) edges are eliminated - this is illustrated
in Fig \ref{mr} where the original tuple graph is split into two sub-graphs neither of wich contains the edge.The idea
is to split a tuple graph recursively in this manner till (i) No sub-graph contains any edges, or (ii) The number of 
sub-graphs exceeds the number of available rows per tuple - whichever is earlier. Each sub-graph then corresponds to a
row in the multi-row representation of the tuple. Consider an uncertain tuple T and three of its attributes $A_i$, $A_j$, and Am with attribute value instances as shown in fig \ref{fig6}. Focusing on attributes $A_i$ and $A_j$, certain attribute value instances in $A_i$ may be inconsistent with certain instances in $A_j$, based on 1 or more (binary) tuple level constraints. For an attribute value instance $a_{i_{k}}$ $\in$ $A_i$, define $cons(a_{i_{k}}, A_j)$ as the set of those attribute value instances in $A_j$ that are consistent with $a_{i_{k}}$ i.e., wrt the tuple level constraints. Now consider a particular row in a multi-row representation for T. A row is said to be consistent wrt attributes $A_i$ and $A_j$ iff all attribute value instances for attribute $A_i$ in that row are consistent with all attribute value instances for $A_j$ in that row. A row is said to be completely consistent if it is consistent wrt all pairs of attributes in the uncertain tuple. A row is inconsistent if it is not consistent wrt at least one pair of attributes $A_i$ and $A_j$. It follows that any row will be inconsistent iff there are 2 attributes $A_i$ and $A_j$ such that there are two instances in $A_i$, $a_{i_{k1}}$ and $a_{i_{k2}}$ and $cons(a_{i_{k1}},Aj) \neq cons(a_{i_{k2}},Aj)$. We denote any such pair of attribute value instances and pair of attributes where this an inconsistency as a {\em violation set} $v=<a_{i_{k1}}, a_{i_{k2}}, Aj> $.

To eliminate the inconsistencies across $A_i$ and $A_j$, the strategy we follow, in the single-row model, is to eliminate certain attribute value instances from attributes $A_i$ and/or $A_j$ till consistency is achieved. This comes at a cost of possibly eliminating certain consistent instances as well and we provided a mechanism to estimate this loss using marginal values in the previous section. We denote as $loss(v)$ the estimate of the consistent mass loss associated with making violation set $v$ consistent by eliminating attribute value instances. 
With the luxury of multiple row, we can instead split a row with a violation set into {\em 2} rows as shown in fig \ref{fig6}. The resulting 2 rows are necessarily consistent wrt $A_i$ and $A_j$. Denote this operation as that of splitting on a violation set. Note that the inconsistencies are eliminated but no consistent mass is lost in the process. We can perform such splitting on all violation sets for the uncertain tuple, the number of violation sets is polynomial in the (maximum) number of attribute value instances in each attribute and the number of attributes. While this will ensure that we end up in a multi-row representation that exactly captures all the consistent tuple instances in the original uncertain tuple, the number of rows created can be exponential. The number of rows however is bounded. Prioritizing and considering violation sets, based on decreasing order of $loss(v)$, we split them till either all inconsistencies are eliminated or we reach the limit of the number of rows, whichever is earlier. Should the limit on number of rows be reached first there will be rows that do have inconsistencies (still) present. We employ the single-row approximation on each of these rows. The heuristic rationale is that the additional row created due to splitting is in a sense saving us the associated loss value of consistent mass.

\section{Integrity Constraint Selection}
In the previous section we studied how individual ICs of different kinds can be applied to remove
inconsistency in an uncertain relation with constraints. Strictly speaking, when we state we are
resolving a tuple (relation) IC we mean we are resolving that tuple (relation) IC in a particular
tuple (NEED-FIX) class that is inconsistent with that IC. This is what the term "resolving an IC"
will imply now on. In this section we describe how a set of ICs can be applied so as to achieve an
approximation $U^\prime$ of good quality. Note that if our goal was to simple eliminate all the
inconsistency we could apply all the ICs and achieve this, however we realize that a significant
amount of consistent instances can be lost this way.  The challenge is to find an optimal {\em
subset} of ICs to apply such that the quality of the approximation achieved is maximized. 

\subsection{Utility of Each IC}
For each {\em individual} IC we can determine whether resolving it will cause the overall quality
to increase or decrease. Assume that for any IC we have an estimate of the inconsistent mass lost,
$IC_{L}$, and the consistent mass lost, $CM_{L}$, as a result of applying that IC. We define
the {\em utility} of an IC, $UT$, as $ UT = IC_{L} -  CM_{L}/Pc$, where $Pc$ is the total
consistent mass in the uncertain relation. The reader can verify, given the quality measure definition in Section 2, that the overall quality will
necessarily increase after resolving that IC if its utility $UT$ is $> 0$. 

We need to be able to determine the utility for any IC. Recall that a tuple inconsistent with a
tuple level IC can be resolved by dropping a single attribute value in some attribute (involved in
the IC violation). $IC_{L}$ in this case can be determined by computing the probability of the
tuple instances in the tuple that are indeed inconsistent w.r.t. that IC. $CM_{L}$ on the other hand is nothing but the
marginal probability of the attribute value instance that we will drop. Determining $UT$ for 
a relation level IC is relatively more complicated.  Recall that resolving a tuple IC in
each NEED-FIX class is a process of eliminating attribute values in possibly multiple tuples.
$IC_{L}$ is determined by statistical sampling within a NEED-FIX class i.e., by randomly
generating relation possibilities from the NEED-FIX class and estimating what is inconsistent .
Now to fix the class if the attribute values to be dropped (across different tuples) are
$av_{1},\cdots, av_{n}$ then $P(av_{1} \cup \cdots \cup av_{n})$ is a measure of the consistent
mass lost by dropping these attributes. This is essentially the estimation of a DNF formula which
can be also be done using Monte-Carlo sampling and applying the Luby-Karp-Madras estimation algorithm \cite{lkm}.

\begin{tabular}{p{7.5cm}}
\\ \hline
\end{tabular}
\begin{tabbing}
{\bf Algorithm: Greedy\_IC\_Resolution} \\
{\em Input:} Uncertain relation $U$, Set of ICs $C$, Threshold $B$ \\
{\em Output:} Sub-relation $U^\prime$ \\
1: $C_t$  $\leftarrow$ $C$ \\
2: $U_t$  $\leftarrow$ $U$ \\
3: $c_m$  $\leftarrow$ $null$ \\
4: initialize\_utilities($C_t$) \\
5: {\bf while} \= ($C_r(U_t)$ $>$ $B$ and $C_t$ $\neq$ $\o$)  \\
6: {\bf do} \\
7: \> UPDATE\_UTILITIES($U_t$,$C_t$) \\
8: \> $c_m$  $\leftarrow$ select\_best\_IC($C_t$) \\
9: \> $U_t$  $\leftarrow$ resolve($U_t$, $c_m$) \\
10: \> $C_t$  $\leftarrow$ $C_t$ - $c_m$\\
11: {\bf end}\\
12: return $U_t$\\
\\
1: UPDATE\_UTILITIES($U_t$,$C_t$) \\
2: {\bf for } \= (each IC $c_i$ in $C_t$) \\
3: \> $benefit(c_i)$  $\leftarrow$  calculate\_benefit($c_i$, $U_t$) \\
4: \> $cost(c_i)$  $\leftarrow$  calculate\_cost($c_i$, $U_t$) \\
5: \> $utility(c_i)$  $\leftarrow$  benefit($c_i$) - cost($c_i$) \\
{\bf initialize\_utilities:} Define a utility value for each IC \\initialized with unknown\\
{\bf select\_best\_IC:} Select the IC with the maximum utility\\
{\bf resolve:} Resolve given IC in the sub-relation according\\ to its type\\
{\bf calculate\_benefit:}  Calculate benefit of given IC, if \\resolved in the sub-relation\\
{\bf calculate\_cost:}  Calculate cost of given IC, if resolved \\in the sub-relation \\
\end{tabbing}
\begin{tabular}{p{7.5cm}}
\\ \hline
\end{tabular}
    
\subsection{IC  Selection}
Based on the utility,  we
need to determine an optimal set of ICs to choose to arrive at a maximum quality approximation.
The complexity in this problem is caused by the fact that there can
be {\em shared dependencies} amongst the resolution for certain ICs, specifically this happens if
some of the attribute values to be dropped are common across multiple ICs. The utility of
applying a set of multiple ICs thus cannot be determined from the utilities of the individual ICs
alone. The problem of determining a subset of ICs that maximizes the resulting approximation
quality, where costs and benefits may be shared across the ICs can in fact be restated as the
Budgeted Maximum Coverage (BMC) problem \cite{bmc}, which unfortunately is NP-Hard. We
thus provide a heuristic algorithm that attempts to find a subset of ICs to apply such that we
achieve an approximation of high (although not necessarily the highest) quality. 

Our approach is to first consider all tuple level ICs and associated tuples and resolve them (or
not). We then move on to considering relation ICs and associated NEED-FIX classes. Within
each of the two categories of ICs we consider and resolve an IC and an associated tuple or
NEED-FIX class in a {\em greedy} fashion. The algorithm selects ICs (and tuples or NEED-FIX
classes) in descending order of the associated utility. After each iteration, the utilities of each of
the ICs (and tuples or NEED-FIX classes) are recalibrated. This is to factor in the attribute value
instances that have already been dropped as a result of the ICs that have so far been applied. The
algorithm applies ICs sequentially in this manner, recomputes utilities at each iteration, and
terminates when we have no more ICs with an associated utility that is $> 0$. Greedy\_IC\_Resolution
describes this algorithm.
\\
{\em Estimation of Key Quantities} In the above approximation and IC selection algorithms
we require the value $Pc$ - total consistent
mass in an uncertain relation, $Cr$ - consistent mass retained and $Ir$ -
inconsistent mass retained for any approximation $U^{\prime}$. Determining any of these values
is also NP-Hard. We state:\\
{\bf Statement 4:} {\em The derivation of the total consistent mass $\delta$ (or $\gamma$ = 1/$\delta$) factor for a U-relation with constraints is NP-Hard} \\
{\bf Proof:} Given an instance of the FD consistency problem, consider the $\delta$ factor for the uncertain relation U in that problem. A consistent relation instance in the original FD consistency problem is present iff $\delta$, the total consistent mass is $>$0. This implies that if $\delta$ (or $\gamma$) can be determined in polynomial time, then the FD consistency problem can be addressed in polynomial time as well. As the original FD consistency problem is NP-Hard, it follows that determining the $\delta$ (or $\gamma$) factor for a U-relation is also NP-Hard.

We thus resort to statistical sampling to estimate these values
instead. A naive approach however is not applicable in this case. Consider estimating $Pc$ given
an uncertain relation U. We can estimate the {\em average} consistent mass per world instance,
$Pc_{AVG}$, and then multiply this by the number of world instances (which we can compute
directly). We recall Hoeffding's inequality \cite{central} from basic
probability theory which states: \\
{\bf Hoeffding's Inequality:} Let $X_1,X_2,...,X_n$ be iid random variables, while for all
$i$ we have $a_i \leq X_i \leq b_i$, and also let $S = \sum_{i=1}^{n} X_i$. Then we have:  
\begin{equation}
\begin{split}
\label{eqhoeff}
Pr(S-E[S] \geq nt) \leq e^{(-2nt^2)/\sum_{i=1}^{n} (b_i-a_i)^2}
\end{split}
\end{equation}
Or: $Pr(SAvgX-EAvgX \geq t) \leq e^{(-2nt^2)/\sum_{i=1}^{n} (b_i-a_i)^2}$ 
\\
where $SAvgX = (\sum_{i=1}^{n} X_i)/n$ is the sample average and EAvgX is the expected
average of the $X_{i}$s. \\
Treating the mass of a single world instance as a random variable $X_i$ above the sample
average $SAvgX$ is an estimate of $Pc_{AVG}$. The value $t$ is a measure of the error. To
estimate a small quantity such as $Pc_{AVG}$ which for an uncertain relation with 3 attributes, 2 attribute values per attribute, and 100
tuples is itself of the order of $2^{-300}$, to
within say a 10\% error requires $t$ to be accordingly small as well. Plugging such a small value
of t and using 0 and 1 as lower and upper bounds for $X_{i}$ we see that we require an
extremely large number of samples (order of $10^{30}$) $n$ to achieve a probability of 0.9 that
the estimation error is within 10\%. Instead of the average consistent mass we estimate the
{\em ratio}, R, of the consistent mass to the total mass. We define a {\em block} in U (or
$U^{\prime}$) to be any subset of relation instances from the possible world of U (or $U^{\prime}$).
For any such block $B_{i}$ define the quantity: \\ 
$R_{B_i}$ = Total consistent mass in $B_{i}$/Total mass in $B_{i}$ \\ 
We choose block size for a block $B_{i}$ such that $R_{B_i}$ can be computed by
exhaustively enumerating through all instances in that block. The average value of $R_{B_i}$, referred
to as AvgR, is simply $\sum_{i=1}^{N} R_{B_i}/N$.
Unlike $Pc_{AVG}$ or $Cr_{AVG}$, AvgR is in general not such an infinitesimally small
quantity (for instance 0.3 could be a value of AvgR). Thus the number of samples required to
estimate AvgR to within a reasonable accuracy is significantly smaller, for instance a confidence
of 0.9 of estimating this to within 10\% error would require sampling just a few hundred such
blocks (Equation~\eqref{eqhoeff}). Now for both $U$ or an approximation $U^{\prime}$ we can
determine the total mass. For $U$ it is simply 1, and for $U^{\prime}$ we can just compute it given
$U^{\prime}$. Having the total mass, and a reasonable estimate of AvgR we can derive reasonably
accurate estimates of $Pc$, or $Cr$ and $Ir$.

\section{Other Issues}
While we have described the basic approach to resolving various kinds of ICs we would like
to discuss some additional issues related to the representation model and the IC resolution
algorithms.
\\
{\em Model Expressivity} The or-set model we have used is simple and efficient but also limited in expressivity.
With more expressive models we will achieve better quality approximations as this will mean having to drop
less consistent mass. We have begun exploring more expressive models with using a {\em mutli-row} representation
for tuples where a tuple can be represented as multiple rows of or-sets of attributes. This is illustrated in
Fig \ref{fig7} where we note that we can now exactly represent the consistent instances of tuple of Fig \ref{utup}.
Our experimental results also show that we achieve better quality approximations using multiple rows. 
Developing an approach for approximating an uncertain relation with constraints to a more complex 
model such as that based on world set decompositions and "ws-sets" \cite{kochicde08} is indeed
an interesting direction for future work.

\begin{figure}
\begin{tabular}{|l|l|l|l|}
\hline
jim & instructor & training (0.6) & MBA (0.3) 1 \\
\hline
jim & instructor  (0.3) & marketing (0.4) & BA (0.7) 1\\
&  &  & MBA (0.3)\\
\hline
\end{tabular}
\caption{Multi-Row Representation}
\label{fig7}
\end{figure}

{\em Incrementality} While in most applications we expect the complete
uncertain relation and set of ICs to be provided upfront, we can also envision 
scenarios where the additions to the ICs, to the uncertain relation itself (i.e., new tuples),
or both, are provided incrementally. Rather than recompute $U^{\prime}$ from scratch in such cases, we present
an incremental approach.  Consider first the case where we have approximated
an uncertain relation $U$ to $U^{\prime}$ given a set of ICs, and are now given a new 
set off additional ICs $C_{Na} \cup C_{Nt} \cup C_{Nr}$, where $C_{Na}$, $C_{Nt}$,
$C_{Nr}$ are the additional attribute, tuple, and relation level ICs respectively. Our approach
is to start with $U^{\prime}$, apply the additional attribute ICs, and then apply
the additional tuple and relation ICs in a greedy fashion using the algorithm
GREEDY\_IC\_Resolution. The steps are as follows:
\\
1. Resolve $C_{Na}$ in $U^{\prime}$ resulting in $U_{1}\prime$\\
2. Resolve $C_{Nt} \cup C_{Nr}$ in a greedy fashion on $U_{1}\prime$, resulting in $U_{2}\prime$ which is now
the new approximation of $U$.  \\
\\
The other case is when new tuples are provided for $U$. Let the 
set of new tuples be $U_{N}$. As attribute and tuple level ICs are local
to individual tuples we need resolve the (existing) attribute and tuple
level ICs {\em only} in $U_{N}$. New violations of relation level ICs
however can occur within the tuples in $U_{N}$ or across the tuple in $U_{N}$
and $U^{\prime}$. We thus proceed as follows. \\ 
1. Resolve $C_{Na}$ in $U_{N}$ resulting in $U_{N1}$. \\
2. Resolve $C_{Nt}$ in a greedy fashion on $U_{N1}$ resulting in $U_{N2}$. \\
3. Resolve $C_{Nr}$ in a greedy fashion on $U\prime \cup U_{N2}$, resulting in $U_{1}\prime$  which is now the new approximation
of $U$.\\
\\
{\em Operations and ICs}
We achieve consistency with the ICs by essentially deleting tuples (the deletion
of an attribute value instance can be viewed as deleting the tuple instances that
get dropped as a consequence). In database {\em repair} one can in general consider
any of tuple deletion, addition, or modification to repair a database to make
it consistent with a set of given ICs.  Our model
is to start with a {\em complete} uncertain relation i.e., one where we know
of {\em all} the possible relations that that uncertain relation implies. Starting
with the complete space of possibilities, the only meaningful operation to ensure
consistency given ICs, is to eliminate possible relations that are inconsistent
with any of the ICs. Coming back to a repair perspective, the deletion of 
tuples is the only viable option in this framework. Another related aspect is that we permit only particular
subtypes of ICs within the classes of relation ICs addressed as shown in Table II.
This is to ensure that a NEED-FIX  class wrt these kinds of constraints can {\em always} be fixed using tuple deletion.

\section{Experimental Evaluation}

We present experimental evaluation results in two different experimental set-ups. The first set-up is to assess the impact of incorporating constraints on applications that use uncertain relations \-- specifically we choose the application of information extraction, and assess an eventual improvement in extraction accuracy with the use of constraints. This experiment is over a real dataset of free text bios of researchers collected from their homepages on the open Web. The second set-up is to evaluate the effectiveness of our approach for approximating an uncertain relation with constraints and our primary goal is to assess the quality of the approximations achieved. We employ a synthetic dataset in this case. We describe below the two sets of experiments and results.

\subsection{Application Impact}
We consider the application of information extraction (IE), in particular the task of "slot-filling" or extracting relations from text. Our goal is to assess any improvement in extraction accuracy that can be achieved with the use of ICs. We store the extracted data provided by a given extractor in an uncertain relation. We further define a set of ICs that are meaningful for the particular relation that is being extracted. We then compare the accuracy of retrieval done over the original uncertain relation, with the uncertain relation refined incorporating the ICs. 

\subsubsection{Dataset}
We have chosen the extraction task of extracting details of a researcher, such as her job-title, employer, academic degrees and their associated dates and alma-maters from {\em free text} bios on their Web pages. We have collected around 500 such Web pages of bios from the homepages of researchers in the field of computer science. We identified 48 different items or slots to be extracted from each Web page which correspond to the above mentioned data items such as degrees, dates, employers etc.

\subsubsection{Uncertain Relation Representation}
We then trained and employed the TIES \cite{ties} information extraction system to extract these slots from the collection of Web pages. The extracted data is first represented in an uncertain relation. We consider each Web page as providing the data for a single tuple in this relation. State-of-the-art extraction systems such as TIES now provide a {\em space} of multiple possibilities for an extracted value for a slot, typically having each possible value associated with a confidence score. The extracted values provided by the extractor for a particular slot are part of the space of attribute values for the corresponding attribute and tuple in the uncertain relation. Also other possible values for that slot, identified through a {\em tokenization} process are included in the space of possible attribute values, realizing a complete space of attribute value possibilities. 
As an example, for a particular page (tuple) say the extractor returns the set of values [(2005 9.2) (2001 1.3)] for the {\tt PhD Date} attribute i.e., two possible values and associated confidence scores. Also suppose that through tokenization we know that one other token, (2003), which is also of the type date (which is the domain for the {\tt PhD Date} attribute), {\em could also} be a value for that slot. The attribute world formed based on this information is $\{$(2005 0.6), (2001, 0.1), (2003, 0.3) $\}$.  (The details such as the determination of the probabilistic distribution in each attribute world are important in general, but not to this discussion). The set of attribute worlds corresponding to all slots for a page forms an uncertain tuple and the set of all such tuples (corresponding to all pages) corresponds to the extracted uncertain relation  that we will call $U_{bios}$.

\subsubsection{Integrity Constraints}
Next, we author a set of integrity constraints that capture the semantics of the bios relation. For instance we know that people receive their PhD degrees only {\em after} their bachelors degrees (in the same major at least), or we know that a person who received his PhD in 1978 is not likely to have a current job title of an Assistant Professor. For this domain we were able to specify a total of over 40 ICs spanning the attribute, tuple, and relation levels. A subset of such constraints are: 
{\em 1) All computer science degrees were awarded after 1959. 2) A person receives his doctoral degree only after his bachelors degree (same major). 3) A NULL value for a degree implies NULL values for
the associated alma-mater and degree date. 4) The PhD degree alma mater and employer of a person
are different. } The first constraint above can be expressed as an attribute
level IC while the other 3 can be expressed as tuple ICs over $U_{bios}$. Strictly speaking, some of the above
constraints (such as 4) are "soft" constraints in that they hold
mostly but not necessarily always. For our purpose we treat
them as hard constraints.

\subsubsection{Results}
We evaluated the precision and recall of retrieval over several different slots in $U_{bios}$. We compare the accuracy of retrieval over the original extracted uncertain relation $U_{bios}$, with that over $U_{bios}$ augmented with the domain ICs. We consider precision and recall on a per-slot basis, where:\\
{\em Precision} for a slot s, PR(s), is defined as:

\begin{equation}
\label{eq:pr}
PR(S) = \sum_{all\,tuples\,t} p(v)_{s,t}/N
\end{equation}

where v is the correct value for the slot s in tuple t, $p(v)_{s,t}$ is the probability associated with value v for slot s in tuple t, and N is the number of tuples returned. \\
{\em Recall} for a slot s, RE(s), is defined as:

\begin{equation}
\label{eq:rec}
PR(S) = \sum_{all\,tuples\,retrieved\,r} p(v)_{s,r}/ \sum_{all\,tuples\,t} p(v)_{s,t}
\end{equation}

where v is the value for slot s in tuple t.

\begin{table}
\begin{tabular}{|p{2.3cm}|c|c|c|}
\hline
\scriptsize{\bf {Slot}} & $p_i$ $\mid$ $p_c$ & $r_i$ $\mid$ $r_c$ & $f_i$ $\mid$ $f_c$ \\ \hline
\scriptsize{\bf {Title}} & $0.95 \mid 0.8$ & $0.78 \mid 0.94$ & $0.85 \mid 0.82$\\ \hline 
\scriptsize{\bf {Employer}} & $0.79 \mid 0.82$ & $0.65 \mid 0.69$ & $0.71 \mid 0.75$\\ \hline 
\scriptsize{\bf {PhD Degree}} & $0.98 \mid 0.98$ & $1 \mid 1$ & $0.98 \mid 0.98$ \\ \hline 
\scriptsize{\bf {PhD School}} & $0.69 \mid 0.76$ & $0.36 \mid 0.58$ & $0.47 \mid 0.66$ \\ \hline 
\scriptsize{\bf {PhD Date}} & $0.69 \mid 0.86$ & $0.46 \mid 0.83$ & $0.55 \mid 0.84$\\ \hline 
\scriptsize{\bf {Bach School}} & $0.93 \mid 0.9$ & $0.3 \mid 0.49$ & $0.45 \mid 0.63$\\ \hline 
\scriptsize{\bf {Bach Date}} & $0.88 \mid 1$ & $0.62 \mid 0.96$ & $0.73 \mid 0.98$\\ \hline 
\end{tabular}
\label{tab3}
\caption{Extraction Accuracy with Constraints}
\end{table}

\begin{figure}[t]
\begin{center}
\includegraphics[width=2.7in]{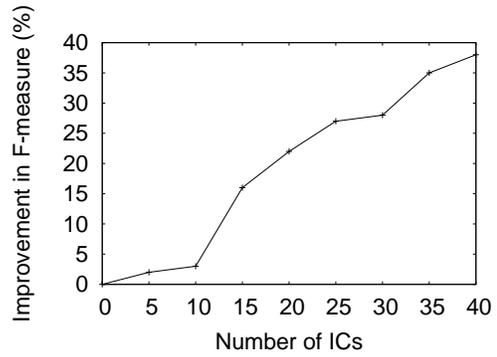}
\end{center}
\caption{Extraction Accuracy}
\label{FM}
\end{figure}

Given $U_{bios}$ and the set of ICs specified over this relation we generate an approximation of $U_{bios}$  plus the ICs, $U_{bios}\prime$ using our approach.  Table 3 provides the retrieval accuracy, in terms of precision, recall, and f-measure, for a subset (for brevity) of the slots over both $U_{bios}$ and $U_{bios}\prime$. Here $p_{I}$, $r_{I}$, and $f_{I}$ are precision, recall and f-measure respectively over $U_{bios}$,  and $p_{c}$, $r_{c}$, and $f_{c}$ are the corresponding values over $U_{bios}\prime$ . We observe that both precision and recall for many slots are significantly improved in $U_{bios}\prime$ compared to $U_{bios}$, thus demonstrating the effectiveness of employing ICs. Albeit in some cases we see a (minor) drop in precision which is due to treating what should be soft constraints as hard. Note that these are extraction accuracy improvements over the output of extraction systems that are representative of the state of the art and also have been provided extensive training data in the application domain. These results demonstrate the utility of employing constraints in the context of an actual application of information extraction where the use of constraints significantly improves the retrieval quality. We also demonstrate (in Fig \ref{FM}) the increase in overall extraction accuracy (aggregated over all the slots in the relation) as a function of the number of ICs incorporated. 

\subsection{Assessing Effectiveness of Approximation Approach}
Our aim is to arrive at a good quality approximation of an uncertain relation with constraints. For more detailed analysis of scalability, sensitivity, and robustness of our algorithm we performed empirical evaluation on synthetic data. We evaluate the quality of approximation that we can achieve with our greedy algorithm. We also compare our results with the alternative algorithm of removing {\em all} inconsistent instances.

\subsubsection{Synthetic Dataset Generation}
We implemented an uncertain relation generator which lets us generate uncertain relations under pre-specified settings for different parameters. The key parameters are described in Fig ~\ref{param} below. The generation parameters allow us to control and configure various factors such as the size of the uncertain relation, uncertainty in the uncertain relation, kinds and number of ICs, the "dirtiness" of the relation i.e., the degree of inconsistency in the original relation etc.

\begin{figure}
\begin{tabular}{|p{0.8cm}|p{7.2cm}|}
\hline
{\bf Param} & {\bf Description}\\ \hline
A & Number of attributes in relation.\\ \hline
$MAX$ & Maximum number of choices in one attribute\\ \hline
C & Total Number of ICs\\ \hline
D & Maximum arity of a (tuple) IC\\ \hline
R & Number of tuples\\ \hline
$\alpha$ & Degree of data dirtiness (\% fields uncertain) \\ \hline
\end{tabular}
\caption{Synthetic Data Generator Parameters}
\label{param}
\end{figure}

The generation of an uncertain relation with constraints comprises of the following basic steps: 1) Generate an initial (clean) uncertain relation according to the relation size, schema size, and relation uncertainty degree parameters. This includes the definition of a probability distribution over the uncertain relation. 2)Generate specific ICs at attribute, tuple, and relation levels based on the number of ICs parameter. 3) Inject instances of violations for the attribute, tuple, and relation level ICs in randomly chosen attributes, tuples, and sets of tuples (respectively) according to the degree of dirtiness parameters. 

\subsubsection{Experimental Results}
\begin{figure}
\begin{center}
	\includegraphics[width=2.7in]{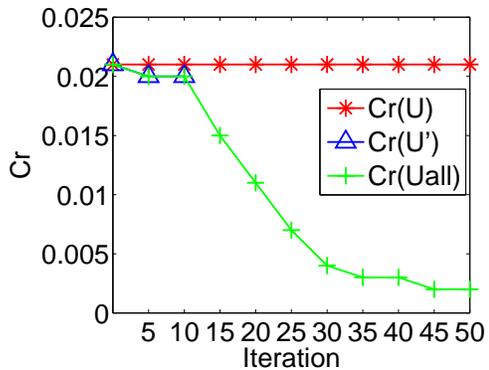}
\\
(a) Consistent Mass Retained
	
	\includegraphics[width=2.7in]{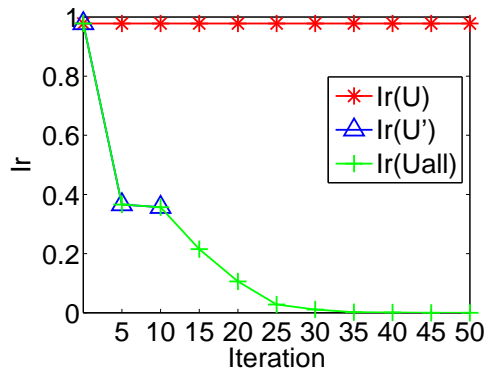}\\
	(b) Inconsistent Mass Retained
\end{center}
\caption{Cr and Ir in each iteration}
\label{crir}
\end{figure}

\begin{figure}
\begin{center}
	\includegraphics[width=2.7in]{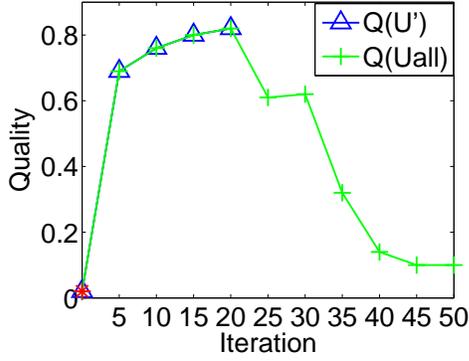}
	\\
(a) Quality

	\includegraphics[width=2.7in]{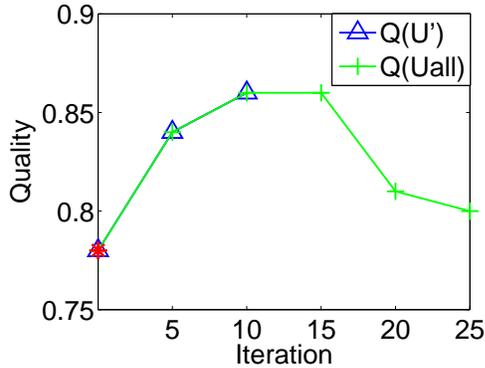}\\
	(b) Quality\\
\end{center}
\caption{Quality after resolving each IC}
\label{QualStep}
\end{figure}

On a synthetic dataset of 1000 tuples with 25 ICs (of different types) Figure \ref{crir}
demonstrates the consistent (Cr) and inconsistent mass (Ir) in the approximation as a function 
of the number of ICs (iterations) applied. Figure \ref{QualStep} illustrates (for 2 cases
of different initial consistency) the approximation quality
as a function of the IC iterations applied. We applied the ICs in order that the greedy algorithm
selects them, the greedy algorithm terminates according to the utility based criterion whereas
the brute force algorithm of resolving all ICs runs on.
These results are typical of the many traces
we conducted. We clearly see the superiority of our greedy IC selection algorithm
($U^{\prime}$) which terminates when resolving ICs is no longer beneficial, as opposed to the brute
force approach ($U_{all}$) of resolving all ICs that can cause the quality to significantly degenerate.

\begin{figure}
\begin{center}
	\includegraphics[width=2.7in]{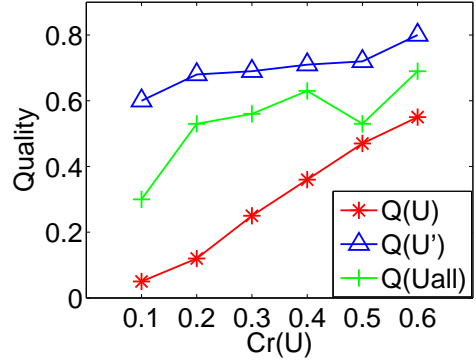}\\
(a)
	
	\includegraphics[width=2.7in]{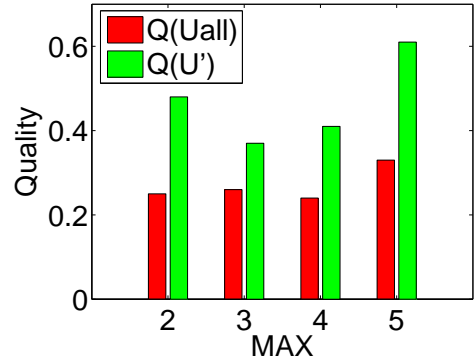}\\
	(b) \\
\end{center}
\caption{Sensitivity}
\label{qvspc}
\end{figure}

In Figure \ref{qvspc} we illustrate the sensitivity of approximation quality (shown averaged
over several traces) to (a) the initial consistent mass in the relation, and (b) the degree of uncertainty
in the original uncertain relation \-- which is controlled by the $MAX$ parameter. We observe (a) that uncertain relations of higher original consistency result in better quality approximations, whereas (b) quality depends
on other factors such as the degree of inconsistency, constraint distribution etc., as opposed to relation uncertainty defined in terms of the number of attribute value choices  $MAX$.

\begin{figure}[t]
\begin{center}
\includegraphics[width=2.9in]{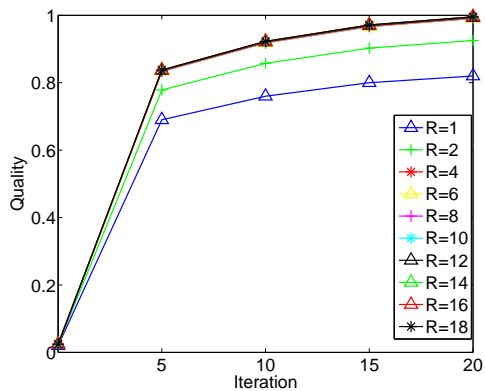}
\end{center}
\caption{Quality in Multi-row Model}
\label{murw}
\end{figure}

\begin{table}[t*]
\begin{center}
\begin{tabular}{|p{1cm}|p{1cm}|p{1.5cm}|p{3cm}|} 
\hline
\# Tuples & \# ICs & Marginals (ms) & IC Resolution (ms) \\ \hline
100 &  5 & $<$ 1 & 703\\ \hline
1000 &  50 & $<$ 1 & 4922 \\ \hline
10000 & 500 & 48 & 99167 (~ 2 min)\\ \hline
50000& 2500 & 1078 & 2591384(~ 53 min)\\ \hline
\end{tabular}
\end{center}
\caption{Time vs Relation Size}
\label{time}
\end{table}

Figure \ref{murw} shows the advantage of using a richer multi\-row model where we can see that the appxroximation quality
increases as more rows are provided for a multi\-row representation of each tuple. Finally in Table \ref{time} we present the 
time required for approximation generation with increasing tuples and IC violations, where we show the time for marginals
computation and the (total) IC resolution time. Note that once the approximation has been generated we can answer queries
very efficiently on the resulting approximation as the constraints have been factored in. The approximation generation
times show that our approach is scalable to large datasets. The experients were conducted on an IBM XSeries\_445 machine
with 4 Intel Xeon 3 GHz processors, 17GB Ram running Windows Server 2003. We must also mention that we have been unable to
provide comparative experimental results with a related system such as MayBMS (in particular) as the "assert" operation
meant to materialize a database recalibrated given an IC is not provided in the current system.

\section{Related Work}
Probabilistic databases have been an area of activity since the 1980s with foundational works such as \cite{barbara, cavallo}
extending the relational model and algebra to represent and support uncertainty in databases.
Current active projects - MystiQ\cite{dalvi07}, MayBMS, Trio, or Orion \cite{tutorial} employ different underlying uncertain database
representation formalisms that either vary subtly, or in some cases significantly across each other, for instance MystiQ
using "or-tuples", Trio using or-sets but with additional "lineage" information, and MayBMS using more expressive
world set decompositions (WSDs). MayBMS has considers conditioning probabilistic 
databases with ICs which is motivated from a data cleaning perspective, dealing with "equality
generating dependencies" (equivalent to the tuple level ICs) and just functional dependencies (FDs)
from amongst relation level ICs (as opposed to the larger class of relation ICs that we address).  
Their approach to resolving ICs is quite different from ours. Instead of applying ICs to an uncertain database as we
do, they augment queries with the ICs so that the ICs are resolved at query time. 
The approach to factor in FDs using a chase based procedure \cite{kochicde08} can result in
an exponential blow up even with a single FD. Each relation is represented as decomposed
into multiple "components" the product of which yields the entire relation. Each component essentially
contains the values of an attribute or a set of attributes.
Their algorithm is to consider pairs of tuples
violating the FD, take each attribute in the FD and merge the components containing those attributes
for the pair of tuples into a new component, and then clean the new component by eliminating 
attribute value combinations that are inconsistent with the FD. In the case of a relation R,
with FD $A \rightarrow B$, and pair-wise violations $(t_1,t_2), (t_2,t_3),,(t_{K-1},t_K)$ with this
FD, we will end up with a component that has as columns ($t_{1}.A, t_{2}.A,..,t_{K}.A, t_{1}.B,..,t_{K}.B$) and
in the rows of this component have all consistent combinations of attribute A and B values. 
The size of this component is $O(M^{K})$ where M is the degree
of uncertainty (choices) in the attributes. Further, the chase based procedure must select the consistent
combinations only and its compelxity is also $O(M^{K})$. Even with modest values of say M=2 and K= 30, $M^{K}$ is extremely large.
While we observe that their approach is exponential, we note that the authors essentially meant the technique to be used in the context of data that has only 
{\em very few} violations, in which case their approach will work fine. This is substantiated by their experiments
which have been done with a degree of data dirtiness as low as 0.001\% - 0.005\% and also stated as a valid
assumption by them given the focus on data cleaning applications. In contrast, our approach is applicable to
databases with a much higher degree of data dirtiness, for applications such as information extraction 
where literally {\em all} fields in the data can be uncertain i.e., with a degree of dirtiness of 100\% !
Also in our synthetic data evaluation we have used an $\alpha$ (dirtiness) factor of at least 5\% (Table \ref{time}).
To the best of our knowledge our work is the first to  a) Provide an approach to factoring a large class of ICs, including many kinds of relation level ICs
such as FDs, aggregation constraints, inclusion dependencies, and set constraints in a correct manner into an uncertain database, b) Provide an approach 
to incorporating ICs that makes no assumptions on factors such as the degree of data dirtiness and is thus applicable to applications where the degree
of data dirtiness can in practice be quite high. 

In information extraction, the approach developed in \cite{sunita} is to approximate a complex CRF distribution that represents text segmentation possibilities into a probabilistic relational model. This work however does not 
consider any dependencies {\em across} different extracted segments, where each extracted segment is treated as a tuple. 
We address such dependencies as relation level ICs.  In \cite{sunita} the probability distribution being
approximated is {\em known} to be generated from a CRF and an efficient
forward-backward-message-passing algorithm is employed for marginal computation, vs our setting where marginal probabilities must be estimated.
We compared with database repair \cite{ lopatenko,bohannon,chomicki07} earlier and further note that most prior work on repair has considered only a limited set of constraints, such as \cite{bohannon} which deals with only functional (FD) and inclusion (IND) dependencies whereas our paper addresses a large class of attribute, tuple, and relation level ICs.  Work
on consistent query answering (CQA) deals with a related but different problem of answering queries over a dirty database considering constraints
over the database - this is established as a hard problem in general \cite{chomicki07} with practical approaches \cite{fuxman05} provided
considering only primary key constraints.

\section{Conclusion}
We have developed an approach for incorporating integrity constraints into uncertain relations by
approximating the uncertain relations. There are several interesting directions for future work, including considering more expressive uncertain database representation models, that we are working on.

\bibliographystyle{IEEEtran}
\bibliography{biblio}

\end{document}